\documentclass[11pt]{article}
\usepackage{amssymb}
\usepackage{amsmath}
\usepackage{graphics}
\usepackage{epsfig}
\usepackage{a4wide}
\usepackage{cite}
\usepackage{multirow}
\usepackage{color}
\usepackage{mathrsfs}

\DeclareGraphicsExtensions{.eps}

\newcommand{\nwa}{\textrm{NWA}}

\begin{document}
\title{ NLO QCD and EW corrections to $WW$+jet production with leptonic $W$-boson decays at LHC}
\author{ Li Wei-Hua$^a$, Zhang Ren-You$^a$, Ma Wen-Gan$^a$, Guo Lei$^b$, Li Xiao-Zhou$^a$, and Zhang Yu$^a$ \\
{\small $^a$ Department of Modern Physics, University of Science and Technology of China, }  \\
{\small $~~$ Hefei, Anhui 230026, P.R.China} \\
{\small $^b$ Department of Physics, Chongqing University, Chongqing 401331, P.R. China} }

\maketitle \vskip 15mm
\begin{abstract}
We present the complete next-to-leading order (NLO) QCD and NLO electroweak (EW) corrections to the $W$-boson pair production associated with a hard jet at the LHC including the spin correlated  $W$-boson leptonic decays. The dependence of the leading order (LO) and NLO corrected cross sections on the jet transverse momentum cut is investigated. In dealing with the subsequent resonant $W$-boson decays, we adopt the {\sc MadSpin} method where the spin correlation effect is included. We also provide the LO and NLO corrected distributions of the transverse momenta and rapidities of final products. The results show that the NLO EW correction is significant in high transverse momentum region due to the EW Sudakov effect.
\end{abstract}


\vfill \eject \baselineskip=0.32in

\renewcommand{\theequation}{\arabic{section}.\arabic{equation}}
\renewcommand{\thesection}{\arabic{section}}

\makeatletter      
\@addtoreset{equation}{section}
\makeatother       

\par
\vskip 5mm
\section{Introduction}
\label{sec:intro}
\par
Both the ATLAS and CMS collaborations at the CERN Large Hadron Collider (LHC) have discovered a neutral boson whose properties are compatible with the Standard Model (SM) Higgs boson \cite{ATLAS-2012,CMS-2012}. The next important goal is to understand the nature of the discovered Higgs boson. Meanwhile, to make precision measurement of Higgs properties, the study of signals of Higgs-boson production and their backgrounds is quite essential.

\par
The $VV^{\prime} + {\rm jet}~(V,V'=W,Z)$ productions at the LHC can be used to determine the gauge boson couplings and will allow for a better understanding of the electroweak (EW) symmetry breaking. Their production cross sections are currently known up to the QCD next-to-leading order (NLO). These production processes are basically backgrounds to the precision measurement of the Higgs-boson production which is used to search for physics beyond the SM. Since the precision measurements will be possible at the LHC Run 2, and the enormous luminosity is permitted to produce gauge bosons with large transverse momentum up to TeV scale, the NLO EW corrections to these processes can receive large logarithmic contributions which originate from soft/collinear exchange of EW gauge bosons in loop diagrams. Therefore, the EW correction could be strongly enhanced by the EW Sudakov logarithms. The EW corrections to most of these processes are yet unknown, although they are certainly as important as QCD corrections in some colliding channels.

\par
The NLO QCD corrections to the $W^+W^- + {\rm jet}$ production at the LHC for on-shell $W$-bosons were calculated in Refs.\cite{Ditt:pp2WWjet,Cambell:pp2WWjet,Ditt:prl}. While the NLO EW corrections to this process with leptonic $W$-boson decays have not been investigated so far. It is believed that the calculations for the $VV'+{\rm jet}$ productions with subsequent vector-boson decays including the NLO QCD and NLO EW corrections at the LHC are desired, and one can find that this work is listed in the wishlist (Table 3) of Ref.\cite{LHwishlist}.

\par
In this paper, we present the LO and NLO QCD+EW corrected integrated cross sections and some kinematic distributions for the $pp \to W^+W^- + {\rm jet} + X$ process at the LHC, including the $W$-boson leptonic decays by adopting the naive narrow width approximation (NWA) and the {\sc MadSpin} program. The NLO correction can be divided in ${\cal O}(\alpha^3\alpha_s)$ EW and ${\cal O}(\alpha^2 \alpha_s^2)$ QCD corrections. The paper is organized as follows: In section \ref{sec:calculation}, we provide a general setup of our calculation. In section~\ref{sec:num}, we present the numerical results and discussion for the integrated and differential cross sections at the LO and NLO. Finally, a short summary is given in section \ref{sec:conclusions}.

\vskip 5mm
\section{ Calculation strategy }
\label{sec:calculation}
\par
Since the calculation strategy of the NLO QCD correction to the $W$-boson pair production in association with a hard jet at the LHC has already been provided in Ref.\cite{Ditt:pp2WWjet}, in this section we give mainly the description of the NLO EW calculation.

\par
\subsection{General description }
\label{sec:generalsetup}
\par
We adopt the five-flavor scheme and neglect the masses of $u$-, $d$-, $c$-, $s$-, $b$-quark in both the NLO QCD and NLO EW calculations \cite{Ditt:pp2WWjet}. Since the long lifetime of bottom quark induces that its decay vertex is away from the primary production point with a resolvable distance, it is possible to exclude experimentally these contributions by supposing that the final $b$-quark jet can be tagged in principle. In other words, by employing $b$-tagging technique we can remove the $W^+W^- + b$-${\rm jet}$ events.

\par
In the evaluation of the NLO QCD corrections in the five-flavor scheme, the singularity from final gluon splitting $g\to b\bar b$ would be incorrectly removed by $b$-tagging, which should be originally canceled by that from the $b$-quark loop in the gluon self-energy. We deal with this problem in the way similar to that in Ref.\cite{Ditt:pp2WWjet}. We consider the two merged bottom quarks with $R < 0.5$ as a ``light'' jet by adopting jet algorithm described in next section. If the two bottom quarks are separated as two individual jets (i.e., $R > 0.5$), we abandon this event due to the $b$-tagging.

\par
Since the NLO QCD correction involves the contributions from the real light-quark emission subprocesses $u_i + \bar{d}_j \rightarrow W^+ + W^- + u_{k} + \bar{d}_l$ and $\bar{u}_i + d_j \rightarrow W^+ + W^- + \bar{u}_k + d_l$ ($i=1,2,3$ and $j,k,l=1,2$) which are related to the CKM matrix elements, we shall keep the non-diagonal Cabibbo-Kobayashi-Maskawa (CKM) matrix in the NLO QCD calculation. In this paper, we take the approximation that the CKM matrix has block-diagonal form allowing quark mixing only between the first two generations since the mixing to the third generation is negligible, i.e.,
\begin{eqnarray}
V_{CKM}
=
\Big( \ V_{ij} \ \Big)_{3 \times 3}
=
\left(
\begin{array}{rcc}
     \cos\theta_C \ &  \sin\theta_C \ &  0 \\
    -\sin\theta_C \ &  \cos\theta_C \ &  0 \\
       0~~~ \ &  0 \ &  1
\end{array}
\right).
\end{eqnarray}
As explained in Ref.\cite{Ditt:pp2WWjet}, the LO contribution for the $pp \rightarrow W^+W^-+{\rm jet} + X$ process is not influenced by the explicit entries of the CKM matrix due to the CKM unitarity. Analogously, the EW one-loop amplitude for $0 \rightarrow W^+ W^- + q_i \bar{q}_j + g$ and the tree-level amplitude for $0 \rightarrow W^+ W^- + q_i \bar{q}_j + g + \gamma$ ($q = u,d$) are dependent on the CKM matrix via only $V_{ik} V_{lk}^{*} V_{lm} V_{jm}^{*}$ and $V_{ik} V_{jk}^{*}$. These combinations of CKM matrix elements equal to $\delta_{ij}$ after the summation over the intermediate quark flavors due to the CKM unitarity. Therefore, the full NLO EW correction to the $pp \rightarrow W^+W^-+{\rm jet}+X$ process is independent of the CKM matrix elements, and we set the CKM matrix to the identity in both the LO and NLO EW calculations.

\par
\begin{figure}[ht!]
\centering
\includegraphics[scale=1]{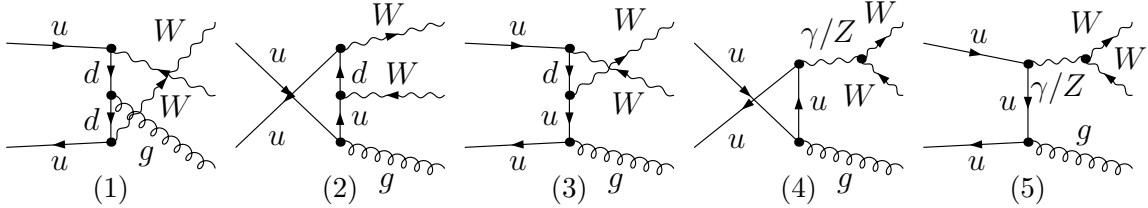}
\caption{\small The LO Feynman diagrams for the partonic process $u\bar{u}\to W^+W^-g$.}
\label{fig:uuwwg}
\end{figure}

The $pp \to W^+W^-+{\rm jet}+X$ production at the LO involves the following partonic processes, (see Fig.\ref{fig:uuwwg} for example)
\begin{eqnarray}
\label{eqn:partonlo1}
q\bar{q} & \to & W^+ W^- g, \\
\label{eqn:partonlo2}
gq & \to & W^+W^- q, \\
g\bar{q} & \to & W^+W^- \bar{q},
\label{eqn:partonlo3} \\
\label{eqn:partonlo4}
b\bar{b} & \to & W^+ W^- g.
\end{eqnarray}
In the partonic processes (\ref{eqn:partonlo1})-(\ref{eqn:partonlo3}), $q$ stands for the light quarks (i.e., $q = u,d,c,s$). Because of the smallness of the (anti)bottom-quark parton distribution function (PDF) in the proton, the contributions of the subprocesses with initial $b$-quark, $gb \to W^+W^- b$, $g\bar{b} \to W^+W^- \bar{b}$ and $b\bar{b} \to W^+ W^- g$, to the hadronic cross section are suppressed with respect to the subprocesses (\ref{eqn:partonlo1})-(\ref{eqn:partonlo3}). As an example, the contribution to the LO total hardronic cross section from the subprocess $b\bar b \to W^+ W^- g$ at  the $14~{\rm TeV}$ LHC is less than $0.26\%$. Therefore, it is reasonable to neglect the NLO EW corrections from the subprocesses involving initial $b$-quark in our calculation. Since we perform the NLO EW calculation for the $W^+W^-+{\rm jet}$ production at the LHC in a precision up to ${\cal O}(\alpha^3 \alpha_s)$, the lowest order contributions from the partonic processes in photon-quark collision and their NLO QCD corrections, which are of the ${\cal O}(\alpha^3)$ and ${\cal O}(\alpha^3 \alpha_s)$, respectively, should be included. These partonic processes, called as the subprocesses with initial photon, are written as
\begin{eqnarray}
\label{eqn:gammalo1}
\gamma q(\bar{q}) & \to&  W^+ W^- q(\bar{q}) \, , ~~~(q=u,d,c,s),
\end{eqnarray}
where the contributions from $\gamma b \to W^+ W^- b$ and $\gamma \bar{b} \to W^+ W^- \bar{b}$ are excluded due the $b$-tagging technique.

\par
In the NLO QCD and NLO EW calculations we apply the {\sc FeynArts}-3.7\cite{FeynArts} package to generate the Feynman diagrams and their analytic amplitudes in the 't Hooft-Feynman gauge. The reductions of algebraic amplitudes are performed by adopting the {\sc FormCalc}-7.3 program \cite{FormCalc}. The scalar and tensor one-loop integrals are evaluated numerically by using our developed in-house library based on {\sc LoopTools}-2.8 \cite{FormCalc}, in which we add an option to switch to quadruple precision codes when the numerical instability is detected. We developed also the interface codes to make use of {\sc MadSpin} program \cite{madspin:theory,madspin:program} via Les Houches event file\cite{LHEF1,LHEF2}.

\par
\subsection{EW virtual correction }
\label{sec:virt}
\par
The EW virtual correction to the $pp \to W^+W^-+{\rm jet} + X$ process involves the contributions from the EW one-loop diagrams for the processes (\ref{eqn:partonlo1})-(\ref{eqn:partonlo3}) and the QCD one-loop diagrams for the processes (\ref{eqn:gammalo1}), which are all at the ${\cal O}(\alpha^3 \alpha_s)$. In our calculation, both the ultraviolet (UV) and infrared (IR) divergences are regularized by adopting the dimensional regularization scheme in $d = 4-2\epsilon$ dimensions. We employ the definitions of the relevant renormalization constants presented in Ref.\cite{Denner:EWPhysics}.

\par
We use the on-shell renormalization scheme to renormalize the relevant masses and wave functions \cite{Denner:EWPhysics}. In the $\alpha(0)$-scheme the fine structure constant $\alpha$ is obtained from the coupling of on-shell $e$-$e$-$\gamma$ three-point function in the Thomson limit. The bare electric charge is related to the renormalized electric charge by $e_0 = (1+\delta Z_e) e$. By means of the Ward identity and the on-shell conditions, the electric charge renormalization constant $\delta Z_e$ in the $\alpha(0)$-scheme can be written as \cite{Denner:EWPhysics}
\begin{eqnarray}\label{Log-1}
\delta Z^{\alpha(0)}_e
=
-\frac{1}{2}\delta Z_{AA} - \frac{1}{2} \tan\theta_W \delta Z_{ZA}
=
\frac{1}{2}\frac{\partial \sum^{AA}_T(p^2) }{\partial p^2}\bigg{|}_{p^2\to 0} - \tan\theta_W \frac{\sum^{AZ}_T(0)}{M^2_Z},
\end{eqnarray}
where $\theta_W$ is the weak mixing angle, and $\sum^{ab}_T(p^2)$ denotes the transverse part of the unrenormalized self-energy at four-momentum squared $p^2$. However, in the soft limit $p^2 \to 0$, the photon self-energy induces large logarithmic terms of $\log(m^2_f/\mu^2)$ where $m_f$ is the light fermion mass and $\mu$ is the typical scale of a process. In the $\alpha(0)$-scheme, each external photon leads to a wave-function renormalization constant $\frac{1}{2}\delta Z_{AA}$, which exactly cancels the large logarithms appearing in the corresponding electric charge renormalization constant $\delta Z_e$. While for a LO process whose number of external photons is less than the number of EW couplings, the uncanceled large logarithms have to be absorbed into the running fine structure constant by adopting the $\alpha(M_Z)$- or $G_{\mu}$-scheme. In this work we take the $G_{\mu}$-scheme, in which $\alpha_{G_{\mu}}$ is defined as
\begin{eqnarray}\label{Log-2}
\alpha_{G_{\mu}} = \frac{\sqrt{2}G_{\mu} M^2_W}{\pi} \left(1-\frac{M^2_W}{M^2_Z}\right)
= \alpha(0)(1+\Delta r) \, ,
\label{eqn:Gmu}
\end{eqnarray}
where $G_{\mu}$ is the Fermi constant, and $\Delta r$ is the NLO EW correction to the muon decay. The explicit expression for $\Delta r$ can be written as\cite{muondecay}
\begin{eqnarray}\label{Log-3}
\Delta r
&=&
-\delta Z_{AA}
- \cot^2\theta_W \left( \frac{\sum^{ZZ}_T(M^2_Z)}{M^2_Z} - \frac{\sum^{WW}_T(M^2_W)}{M^2_W} \right)
+ \frac{\sum^{WW}_T(0)-\sum^{WW}_T(M^2_W)}{M^2_W} \nonumber \\
&& + 2 \cot\theta_W \frac{\sum^{AZ}_T(0)}{M^2_Z}
+ \frac{\alpha(0)}{4\pi \sin^2\theta_W} \left( 6 + \frac{7 -4 \sin^2\theta_W}{2 \sin^2\theta_W} \log(\cos^2\theta_W) \right).
\label{eqn:deltar}
\end{eqnarray}
Then the electric charge renormalization constant in the $G_{\mu}$-scheme can be expressed as
\begin{eqnarray}\label{Log-4}
\delta Z^{G_{\mu}}_e = \delta Z^{\alpha(0)}_e - \frac{1}{2}\Delta r .
\end{eqnarray}
From Eqs.(\ref{Log-1}), (\ref{Log-3})-(\ref{Log-4}) we may obtain the analytic expression for $\delta Z^{G_{\mu}}_e$ in which the large logarithmic terms coming from $\delta Z_{AA}$ are canceled obviously. After performing the renormalization procedure, the UV divergences are canceled exactly.
\begin{figure}[ht!]
\centering
\includegraphics[scale=1]{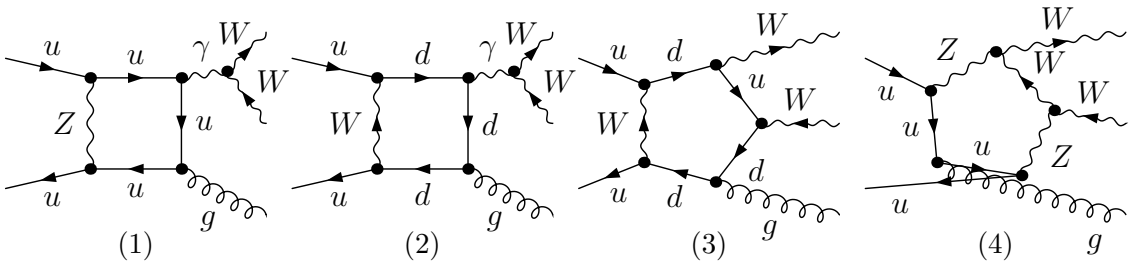}
\caption{\small Some representative box and pentagon diagrams for the partonic process $u\bar{u}\to W^+W^-g$.}
\label{fig:loop}
\end{figure}

\par
Some representative box and pentagon Feynman diagrams for the $u\bar{u}\to W^+W^-g$ partonic process are presented in Fig.\ref{fig:loop}. The amplitude for each one-loop Feynman diagram can be expressed as a one-loop integral. The tensor and scalar 5-point integrals are directly reduced to 4-point integrals \cite{Denner:TensorReduction}. The $N$-point tensor integrals with $N \le 4$ are reduced to scalar integrals with the Passarino-Veltman algorithm \cite{PVreduction}. The above reduction procedures have already been implemented in the {\sc LoopTools}-2.8 package. In the calculation of the 3- and 4-point tensor integrals, we meet the numerical unstability occasionally in the vicinities of the points with zero Gram determinant. To solve this problem in loop integration, we use our developed in-house library which can switch to the quadruple precision codes automatically when small Gram determinant occurs \cite{NLOWWZ,EW:WWphoton}.

\par
\subsection{EW real emission correction }
\label{sec:real}
\par
The real photon emission processes are listed below:
\begin{eqnarray}
\label{eqn:partonlo1real}
q\bar{q} &\to& W^+ W^- g \gamma, \\
\label{eqn:partonlo2real}
gq(\bar{q}) &\to& W^+W^- q(\bar{q}) \gamma \, .
\end{eqnarray}
In calculating real photon emission partonic processes, e.g., $q(p_1)\bar{q}(p_2) \to W^+ (p_3)W^-(p_4) g(p_5) \gamma(p_6)$, we use the two cutoff phase space slicing (TCPSS) method \cite{TCPSS} to isolate the soft and collinear IR singularities. Two arbitrary small cutoffs $\delta_s$ and $\delta_c$ are introduced to separate the phase space into soft ($E_6 \leq \frac{1}{2}\delta_s\sqrt{\hat{s}}$), hard collinear ($E_6  > \frac{1}{2}\delta_s\sqrt{\hat{s}}$, $\hat{s}_{16}$ or $\hat{s}_{26} \leq \delta_c \hat{s}$) and hard noncollinear ($E_6  >\frac{1}{2}\delta_s\sqrt{\hat{s}}$, $\hat{s}_{16}$ and $\hat{s}_{26} > \delta_c \hat{s}$) regions, where $\hat{s}$ is the partonic center-of-mass energy squared and $\hat{s}_{ij} = (p_i + p_j)^2$ is the usual Mandelstam invariant. To get IR-safe results at the LO, we should apply the transverse momentum cut of $p_{T,j}>p_{T,j}^{cut}$ on the leading jet for $W^+W^- + {\rm jet}$ events. For the real emission processes, we merge the two tracks of jets (or jet and photon) into a single track if they are sufficiently collinear. This track combination procedure can lead to collinear IR-safe observables for the process (\ref{eqn:partonlo2real}), but will induce the soft-gluon divergence problem for the process (\ref{eqn:partonlo1real}). This problem can be solved by adopting an event selection criterion (see section~\ref{sec:cuts_selection}) in which $W^+ W^- + {\rm jet}$ and $W^+ W^- + \gamma$ events are properly defined.

\par
\subsection{ Subsequent $W$-boson leptonic decay}
\label{sec:decay}
\par
In the LO evaluation of the $W^+W^-+{\rm jet}$ production followed by $W$-boson leptonic decays, e.g., $pp \to W^+W^-+{\rm jet} \to e^+ \mu^- \nu_e \bar \nu_{\mu}+{\rm jet}+X$, we employ the following two methods to generate the subsequent decays in order to investigate the effect from the resonant $W$-boson spin-entangled decays.
\begin{enumerate}
\item
The naive NWA method. In this method, $W$-boson is treated as an on-shell particle without spin information, and therefore the $W$-boson decay products, lepton and neutrino, can be generated isotropically in the $W$-boson center-of-mass system. The integrated cross section for $pp \to W^+W^-+{\rm jet} \to e^+ \mu^- \nu_e \bar \nu_{\mu}+{\rm jet}+X$ can be easily obtained as
\begin{eqnarray}
&&~~~\sigma\left( pp \to W^+W^-+{\rm jet} \to e^+ \mu^- \nu_e \bar \nu_{\mu}+{\rm jet}+X \right) \nonumber \\
&&=
\sigma\left( pp \to W^+W^-+{\rm jet} +X \right)
\times
Br\left( W^- \rightarrow e^- \bar{\nu}_e \right)
\times
Br\left( W^- \rightarrow \mu^- \bar{\nu}_{\mu} \right),~~~~~~~~~~
\end{eqnarray}
where $Br\left( W^- \rightarrow \ell^- \bar{\nu} \right)$ is the branch ratio for the $W$-boson leptonic decay $W^- \rightarrow \ell^- \bar{\nu}$.
\item
The {\sc MadSpin} method. Since the spin correlation may affect some lepton distributions significantly, the {\sc MadSpin} method taking into account the spin correlation was introduced in Refs.\cite{madspin:theory,madspin:program}. The {\sc MadSpin} program, which is part of {\sc MadGraph5\_aMC@NLO}\cite{mg5}, can be used to generate the heavy resonance decay preserving both spin correlation and finite width effects to a very good accuracy, and is particularly suited for the decay of resonance in production events generated at NLO accuracy \cite{madspin:program}.
\end{enumerate}

\par
In the NLO calculation of the $pp \to W^+W^-+{\rm jet} \to e^+ \mu^- \nu_e \bar \nu_{\mu}+{\rm jet}+X$ process, we shall adopt only the {\sc MadSpin} program to deal with the $W$-boson leptonic decays, which can keep both spin correlation and finite width effects.

\vskip 5mm
\section{Numerical results and discussion}
\label{sec:num}
\par
\subsection{Input parameters}
\label{sec:inputpara}
\par
The relevant SM input parameters are taken as \cite{pdg2014}
\begin{eqnarray}
\label{eq:SMpar}
&& M_W = 80.385~{\rm GeV},~~ M_Z = 91.1876~{\rm GeV},~~ M_H = 125.7~{\rm GeV}, \nonumber \\
&& M_t=173.21~{\rm GeV},~~ G_{\mu} = 1.1663787 \times 10^{-5}~{\rm GeV}^{-2},~~ \alpha_s(M_Z) = 0.118 \, .  ~~~~~~
\end{eqnarray}
In the LO and NLO EW calculations we set the CKM matrix to the identity due to the independence on the CKM matrix elements as described in last section \cite{Ditt:pp2WWjet}. While in the NLO QCD calculations we shall keep the non-diagonal CKM matrix and take the values of the CKM matrix elements as
\begin{eqnarray}\label{CKM}
 V_{CKM} = \left(
\begin{array}{rcc}
     0.97425 \ &  0.22547 \ &  0 \\
    -0.22547 \ &  0.97425 \ &  0 \\
       0~~~~ \ &  0 \ &  1
\end{array}  \right) .
\end{eqnarray}
The $W$-boson decay width is obtained as $\Gamma_W = 2.045294~{\rm GeV}$ by using the {\sc MadSpin} program, and the masses of all leptons and light quarks are set to zero. We take the proton-proton center-of-mass colliding energy to be $14~{\rm TeV}$ and use the NNPDF2.3 QED PDF (NNPDF23-nlo-as-0118-qed set) \cite{NNPDF} in both the LO and NLO EW calculations.

\par
\subsection{Event selection criterion}
\label{sec:cuts_selection}
\par
For the $W^+W^-+{\rm jet}$ production, there exist $W^+W^-+{\rm jet}+\gamma$ four-particle events at the EW NLO. In analyzing these four-particle events, we merge the two tracks of photon and jet into a photon/jet track when $R_{\gamma j} < 0.5$, where $R_{\gamma j} = \sqrt{(y_{\gamma} - y_{j})^2 + \phi^2_{\gamma j}}$. $y_{\gamma}$ and $y_j$ are the rapidities of photon and jet, respectively, and $\phi_{{\gamma j}}$ is the azimuthal-angle difference between the photon and jet on the transverse plane. In the case that the photon and jet are merged, we define a parameter as $z_{\gamma} = E_{\gamma}/(E_{\gamma} + E_j)$ to classify the prototype three-body events \cite{DD:wj,DD:zj}. If $z_{\gamma} > z^{cut}_{\gamma} = 0.7$, the prototype event is called as a $W^+W^-+\gamma$ event and rejected, otherwise, it is treated as a $W^+W^-+{\rm jet}$ event and kept. However, the above event selection scheme leads to the uncanceled collinear IR divergence, because the Kinoshita-Lee-Nauenberg (KLN) theorem \cite{KLNtheorem-K,KLNtheorem-LN} is not violated only when we integrate over the full $z_{\gamma}$ region. To tackle this problem, the bare quark-photon fragmentation function is introduced to absorb the remaining collinear IR divergence. With this method we should subtract the contribution from the range of $z_{\gamma} > z_{\gamma}^{cut}$ , i.e.,
\begin{eqnarray}
d\sigma^{(sub)} = d\sigma_{LO}\int^{1}_{z_{\gamma}^{cut}} {\mathscr D}_{q\to \gamma}(z_{\gamma}) dz_{\gamma} \, ,
\end{eqnarray}
where the effective quark-photon fragmentation function ${\mathscr D}_{q\to \gamma}(z_{\gamma})$ can be expressed as \cite{LEP:ff}
\begin{eqnarray}
\mathscr{D}_{q\to \gamma}(z_{\gamma})  =  -\frac{\alpha Q^2_q}{2\pi}
\frac{1}{\epsilon \Gamma (1-\epsilon)} \left( \frac{4\pi\mu^2_r}{\delta_c s}\right)^{\epsilon}
\left[z_{\gamma}(1-z_{\gamma})\right]^{-\epsilon} \left(P_{q\to\gamma}(z_{\gamma}) - \epsilon z_{\gamma}\right)
+ D^{bare}_{q\to \gamma}(z_{\gamma})
\label{eqn:ff}
\end{eqnarray}
in the dimensional regularization scheme and TCPSS phase space splitting scheme at the EW NLO. The effective quark-photon fragmentation function $\mathscr{D}_{q\to\gamma}(z_{\gamma})$ describes the probability of a quark fragmenting into a jet containing a photon which carries the fraction $z_{\gamma}$ of the total energy. As shown in Eq.(\ref{eqn:ff}), the effective quark-photon fragmentation function consists of two parts, which correspond to the perturbative radiation and non-perturbative production of a photon. At the EW NLO, the bare fragmentation function can be written as \cite{LEP:ff}
\begin{eqnarray}
D^{bare}_{q\to \gamma}(z_{\gamma}) =  \frac{\alpha Q^2_q}{2\pi} \frac{1}{\epsilon \Gamma (1-\epsilon)}
\left(\frac{4\pi\mu_r^2}{\mu^2_f} \right)^2 P_{q\to\gamma}(z_{\gamma}) + D_{q\to\gamma}(z_{\gamma},\mu_f)
\label{eqn:bareff}
\end{eqnarray}
in the $\overline{MS}$ renormalization scheme. Then we obtain
\begin{eqnarray}
{\mathscr D}_{q\to \gamma}(z_{\gamma}) = \frac{\alpha Q^2_q}{2\pi}
\left[P_{q\to\gamma}\left(z_{\gamma}\right) \ln\left(\frac{z_{\gamma}\left(1-z_{\gamma}\right)\delta_c s}{\mu^2_f}\right) + z_{\gamma}\right] + D_{q\to\gamma}(z_{\gamma},\mu_f)\, ,
\end{eqnarray}
where $D_{q\to\gamma}(z_{\gamma}, \mu_f)$ is the finite part of the bare non-perturbative quark-photon fragmentation function. It is parametrized as
\begin{eqnarray}
D_{q\to\gamma}(z_{\gamma}, \mu_f) = \frac{\alpha Q^2_q}{2\pi}\left[ P_{q\to\gamma}(z_{\gamma})
\ln\left( \frac{\mu^2_f}{(1-z_{\gamma})^2\mu^2_0}\right) + C \right],
\end{eqnarray}
and has been measured by the ALEPH collaboration \cite{ALEPH}, where $\mu_0 = 0.14~{\rm GeV}$ and $C = -13.26$, and $P_{q\to\gamma}(z_{\gamma})$ is the quark-to-photon splitting function expressed as
\begin{eqnarray}
P_{q\to\gamma}(z_{\gamma}) = \frac{1+(1-z_{\gamma})^2}{z_{\gamma}} \, .
\label{eqn:qasplit}
\end{eqnarray}

\par
In our NLO EW calculation, we have checked the independence of the total cross section on the two cutoffs $\delta_s$ and $\delta_c$ numerically. For the subprocesses involving two jets in the final state, we merge the two final jets $j_1$ and $j_2$ into one jet when $R_{j_1j_2} < 0.5$. In analyzing the $pp \to W^+W^-+{\rm jet} \to e^+\mu^-\nu_e\bar{\nu}_{\mu}+{\rm jet}+X$ process, we apply the following additional constraints on the final jets and leptons:
\begin{eqnarray}
|y_j| < 4.5,\quad |y_{\ell}| < 2.5, \quad p_{T,\ell} > 25~{\rm GeV},\quad p_{\textrm{T}, {\rm miss}} > 25~{\rm GeV}, \quad R_{j\ell} > 0.4, \quad R_{\ell \ell} > 0.2 \, ,
\label{eqn:finalcut1}
\end{eqnarray}
where $p_{T,\ell}$ is the transverse momentum of the final charged lepton. As pointed out in Ref.\cite{Ditt:pp2WWjet}, the missing transverse momentum is particularly of interest if the $W^+W^-+{\rm jet}$ production is considered as a background process for SUSY searches.

\par
\subsection{Total cross sections}
\label{sec:totalcs}
\par
To confirm the validity of our codes in calculating NLO corrections, we compared our NLO QCD results obtained by employing the {\sc FeynArts}-3.7/{\sc FormCalc}-7.3/{\sc LoopTools}-2.8 programs with the corresponding ones by applying the {\sc MadGraph5\_aMC@NLO} package in the four-flavor scheme, and find that both are coincident with each other. We also used the same scheme and parameters as in Ref.\cite{Ditt:pp2WWjet} and get the NLO QCD corrected integrated cross section as $31.921(35)~pb$, which is in good agreement with the corresponding result as $31.970(11)~pb$ in Ref.\cite{Ditt:pp2WWjet}.

\par
In Table~\ref{tab:ptcut} we list the LO, NLO QCD, NLO EW and NLO QCD+EW corrected cross sections and the corresponding relative corrections to the $W^+W^-+{\rm jet}$ production at the $14~ {\rm TeV}$ LHC separately. The relative NLO QCD, NLO EW and NLO QCD+EW corrections are defined as
\begin{eqnarray}
\delta^{{\rm QCD}} = \frac{\sigma^{{\rm QCD}}_{{\rm NLO}}}{\sigma_{{\rm LO}}} - 1,~~~~
\delta^{{\rm EW}} = \frac{\sigma^{{\rm EW}}_{{\rm NLO}}}{\sigma_{{\rm LO}}} - 1,~~~~
\delta^{{\rm QCD + EW}} = \frac{\sigma^{{\rm QCD + EW}}_{{\rm NLO}}}{\sigma_{{\rm LO}}} -1.
\end{eqnarray}
The factorization and renormalization scales are chosen as $\mu_f = \mu_r = \mu = M_W$ and $\mu_0$ for comparison, where $\mu_0$ is a dynamic scale defined as $\mu_0 = \sqrt{2 M^2_W + \sum_{i}p^2_{T,i}}$ with $i$ running over the final photon and jets. From this table we find that with the increment of $p^{cut}_{T,j}$, the NLO EW correction becomes more significant due to the large EW Sudakov logarithms. For $p^{cut}_{T,j} = 1000~{\rm GeV}$, the NLO EW correction is about $-22.2\%$ and $-26.1\%$ by taking $\mu = M_W$ and $\mu_0$, respectively.
\begin{table}
\center
\begin{tabular}{|c|c|ccccc|}
\hline
\multicolumn{2}{|c|}{$p^{cut}_{T,j}$~[GeV]}   & 50 &  100  &  200 & 500 & 1000   \\
\hline
\multirow{2}{*}{$\sigma_{{\rm LO}}$~[fb]}  & $\mu = M_W$ & 22976(5) & 10035(2)  & 2833.8(6) & 208.56(5) & 10.971(2) \\
&  $\mu = \mu_0$ & 20492(4) &  8407(2)  &  2100.1(5) & 115.25(3) & 4.329(1)   \\
\hline
\multirow{2}{*}{$\sigma^{{\rm QCD}}_{{\rm NLO}}$~[fb]} & $\mu = M_W$ & 30416(67) & 13685(50)  & 4077(15) & 348(1) & 23.48(8) \\
& $\mu = \mu_0$ & 28912(55)  &  13006(39)  &  3747(10) & 268(1) & 13.25(7) \\
\hline
\multirow{2}{*}{$\delta^{{\rm QCD}}$~[\%]} & $\mu = M_W$ & 32.4 & 36.4  & 43.9 & 66.9 & 114 \\
& $\mu = \mu_0$ & 41.1  & 54.7 &  78.4 &  132.5 & 206.1 \\
\hline
\multirow{2}{*}{$\sigma^{{\rm EW}}_{{\rm NLO}}$~[fb]} & $\mu = M_W$ & 23344(11) & 10059(6)  & 2733(2) & 180.8(1) & 8.53(1) \\
& $\mu = \mu_0$ & 20972(7) &  8588(5)  & 2095(2) & 92.3(8) & 3.2(2)  \\
\hline
\multirow{2}{*}{$\delta^{{\rm EW}}$~[\%]} & $\mu = M_W$ & 1.6 & 0.24  & -3.6 & -13.3 & -22.2 \\
& $\mu = \mu_0$ &  2.3  &  2.2  & -0.2  & -19.9 & -26.1 \\
\hline
\multirow{2}{*}{$\sigma^{{\rm QCD + EW}}_{{\rm NLO}}$~[fb]} & $\mu = M_W$ & 30784(68) & 13709(51) & 3976(15) & 320(1) & 21.04(8) \\
& $\mu = \mu_0$ & 29392(53) & 13187(39) & 3742(11) & 245(1) & 12.12(8) \\
\hline
\multirow{2}{*}{$\delta^{{\rm QCD + EW}}$~[\%]} & $\mu = M_W$ & 34.0 & 36.6  & 40.3 & 53.5 & 91.8 \\
& $\mu = \mu_0$ & 43.4 & 56.9 & 78.2 & 112.6 & 180.0 \\
\hline
\end{tabular}
\caption{\small The LO, NLO QCD, NLO EW and NLO QCD+EW corrected integrated cross sections and the corresponding relative corrections to the $W^+W^-+{\rm jet}$ production at the $14~ {\rm TeV}$ LHC by taking $\mu = M_W$ and $\mu_0$ for some typical values of $p^{cut}_{T,j}$.}
\label{tab:ptcut}
\end{table}

\par
In order to explore the theoretical uncertainty from the factorization and renormalization scales, we define the upper and lower relative scale uncertainties as
\begin{eqnarray}
\eta_{+} = \frac{\max\{\sigma(\mu_f, \mu_r)\}}{\sigma(\mu_0, \mu_0)} - 1,~~~~
\eta_{-} = \frac{\min\{\sigma(\mu_f, \mu_r)\}}{\sigma(\mu_0, \mu_0)} - 1,
\end{eqnarray}
where $\mu_f$ and $\mu_r$ run over the following set:
\begin{eqnarray}
\left\{
\left( \frac{\mu_f}{\mu_0}, \frac{\mu_r}{\mu_0} \right)
\biggl| \,
\left( 2, 2 \right), \,
\left( 2, 1/2 \right), \,
\left( 1/2, 2 \right), \,
\left( 1/2, 1/2 \right)
\right\}.
\end{eqnarray}
For $p^{cut}_{T,j}=50~{\rm GeV}$, the LO and NLO QCD+EW corrected integrated cross sections at the $14~{\rm TeV}$ LHC are obtained as $\sigma_{{\rm LO}} = 20492^{+11\%}_{-9.5\%}~fb$ and $\sigma_{{\rm NLO}}^{{\rm QCD + EW}} = 29392^{+4.3\%}_{-5.5\%}~fb$, respectively \footnote{An integrated cross section with upper and lower scale uncertainties is denoted as $\sigma(\mu_0,\mu_0)^{\eta_+}_{\eta_-}$.}. It shows obviously that the relative scale uncertainty can be improved by including the NLO QCD and NLO EW corrections. In further analysis we set the factorization and renormalization scales as $\mu_f = \mu_r = M_W$.

\par
\subsection{Differential cross sections}
\label{sec:diffcs}
\par
In this subsection we study the differential cross sections for the $pp \to W^+W^-+{\rm jet} \to e^+\mu^-\nu_e\bar{\nu}_{\mu}+{\rm jet}+X$ process.

\par
\subsubsection{LO distributions }
\label{sec:lodis}
\par
\begin{figure}[ht!]
\centering
\includegraphics[scale=0.35]{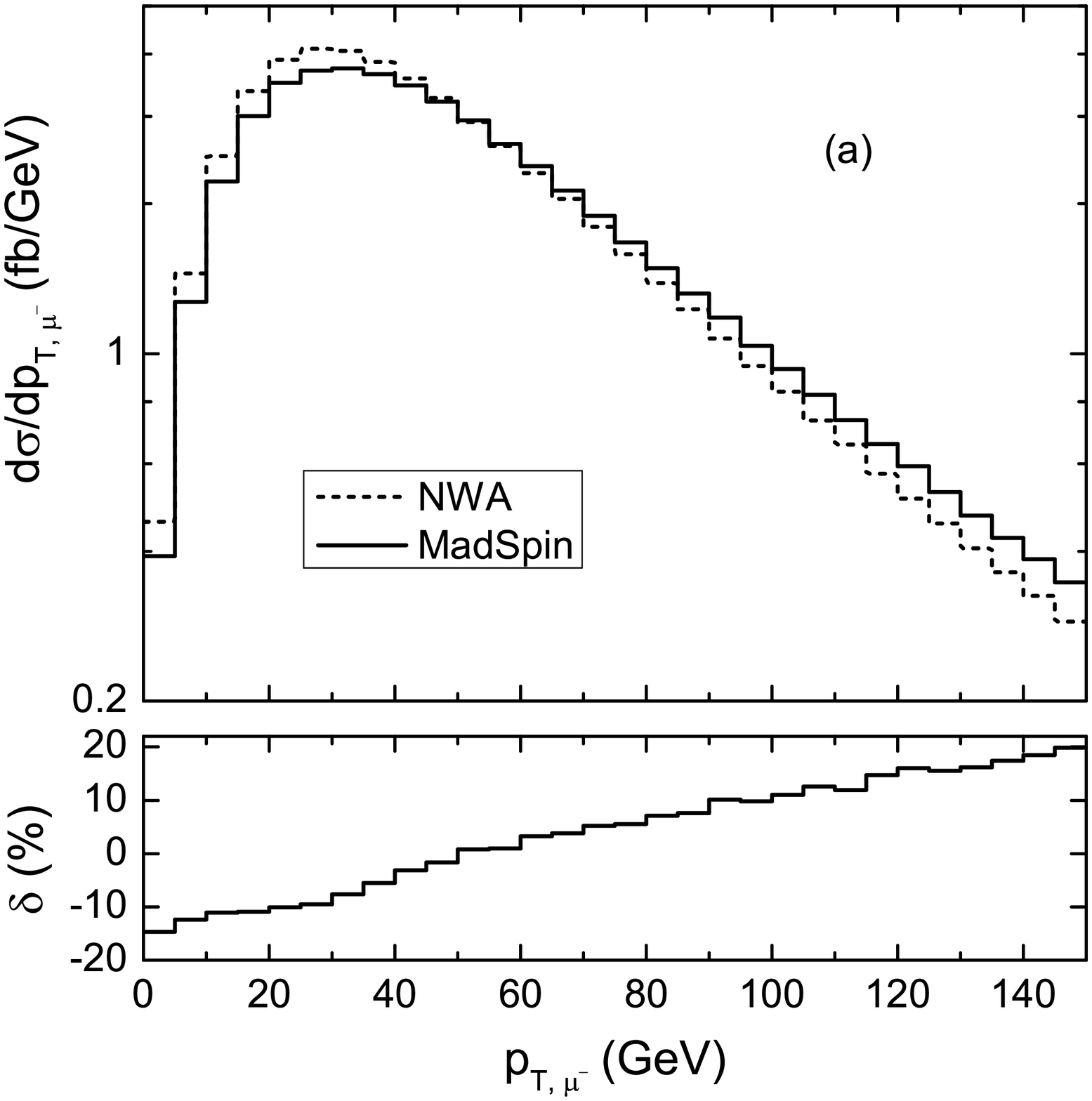}
\includegraphics[scale=0.35]{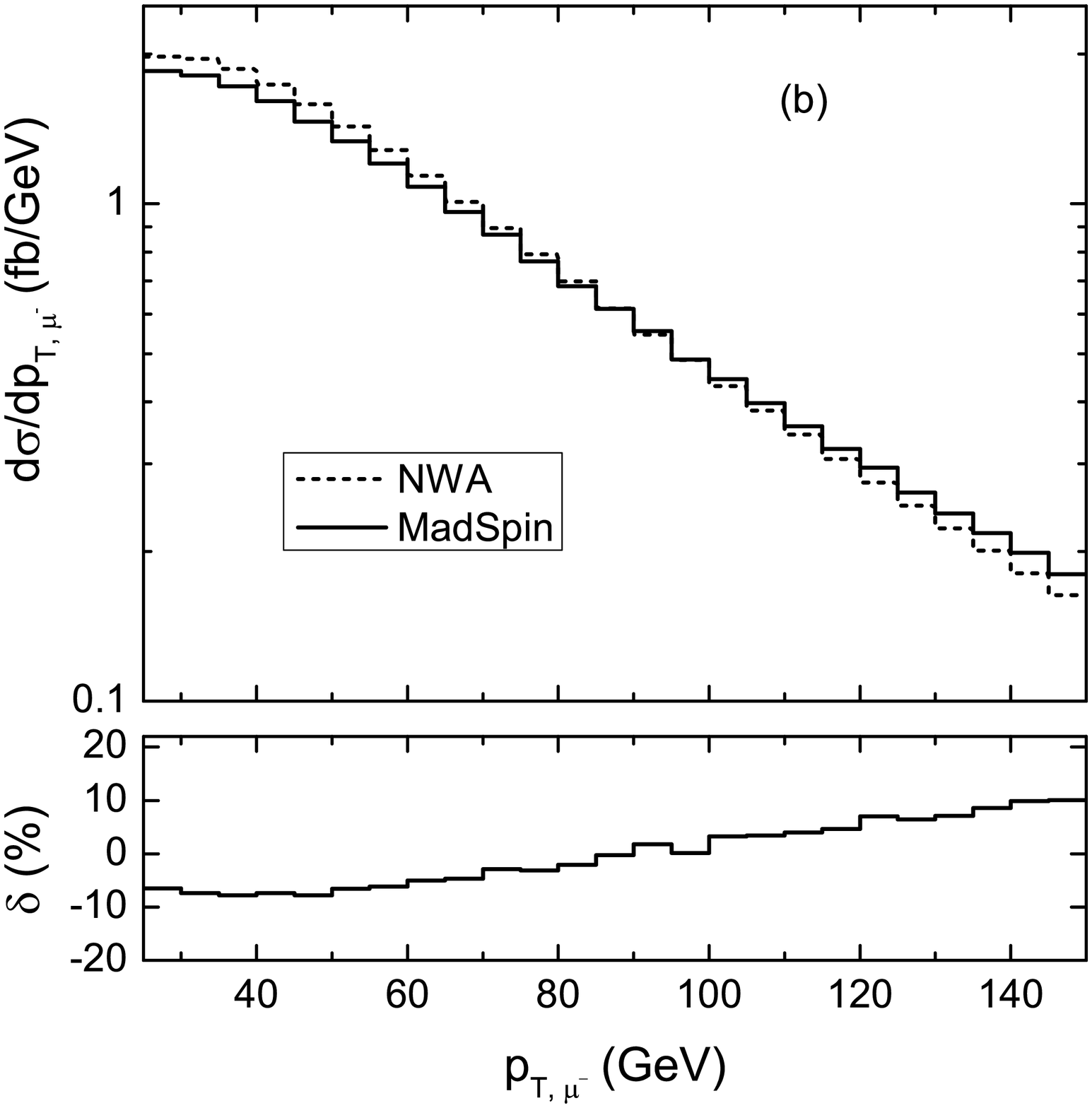}
\caption{\small The LO transverse momentum distributions of $\mu^-$ and the corresponding relative deviations by applying the naive NWA and {\sc MadSpin} methods with (a) $p_{T,j} > 50~{\rm GeV}$, and (b) $p_{T,j} >  50~{\rm GeV}$ and the additional constraints in (\ref{eqn:finalcut1}). }
\label{fig:lomupt}
\end{figure}

We calculate the LO kinematic distributions of the final particles after $W$-boson decays by applying both the naive NWA and {\sc MadSpin} methods as declared in section \ref{sec:decay} in order to display the spin correlation effect. The spin correlation effect in differential cross sections can be acquired by comparing the results obtained by using these two methods, and quantitatively described by the relative deviation defined as
\begin{eqnarray}
\delta(x) = \left( \frac{d\sigma_{\rm MadSpin}}{dx} - \frac{d\sigma_{\nwa}}{dx} \right)\biggl/\frac{d\sigma_{\nwa}}{dx} \, .
\end{eqnarray}
We present the LO transverse momentum distributions of $\mu^-$ and the corresponding relative deviations by adopting two event selection schemes in Figs.\ref{fig:lomupt}(a) and (b) separately. In Fig.\ref{fig:lomupt}(a) we apply only the transverse momentum cut of $p_{T,j} > 50~{\rm GeV}$ on the leading jet. We see from this figure that there exists discrepancy between the two $p_{T,\mu^-}$ distributions obtained by adopting the {\sc MadSpin} and the naive NWA methods separately. The discrepancy is particularly obvious in the low $p_{T,\mu^-}$ ($p_{T,\mu^-} < 30~{\rm GeV}$) and high $p_{T,\mu^-}$ ($p_{T,\mu^-} > 100~{\rm GeV}$) regions, and the relative deviation for the $p_{T,\mu^-}$ distribution can reach $19.8\%$ in the plotted $p_{T,\mu^-}$ range. In Fig.\ref{fig:lomupt}(b) we take not only the leading jet transverse momentum cut of $p_{T,j} > 50~{\rm GeV}$, but also the additional constraints in Eq.(\ref{eqn:finalcut1}) in the event selection, and we see that the relative deviation varies from $-7.8\%$ to $10.1\%$ in the plotted $p_{T,\mu^-}$ region.

\par
\begin{figure}[ht!]
\centering
\includegraphics[scale=0.35]{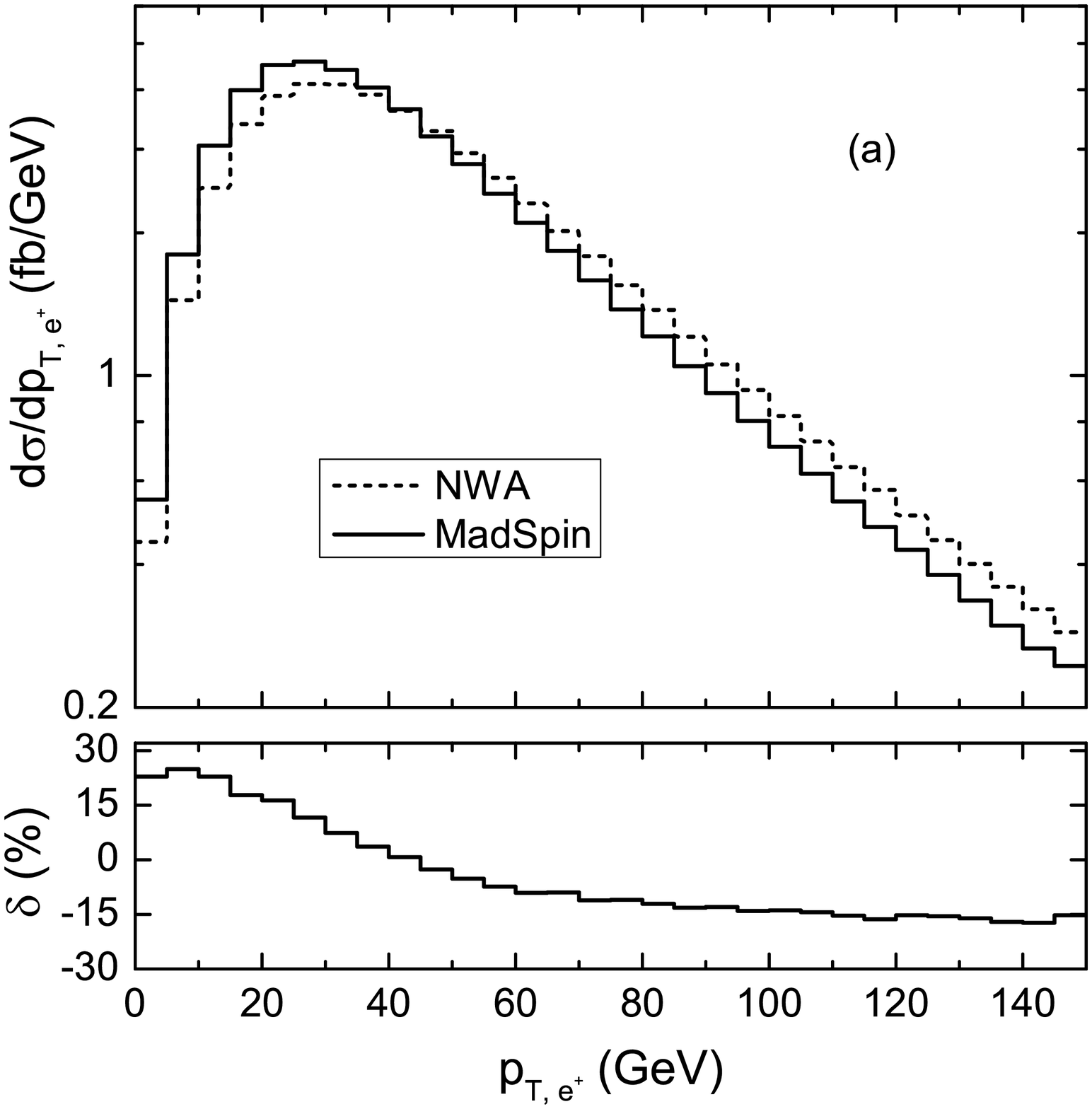}
\includegraphics[scale=0.35]{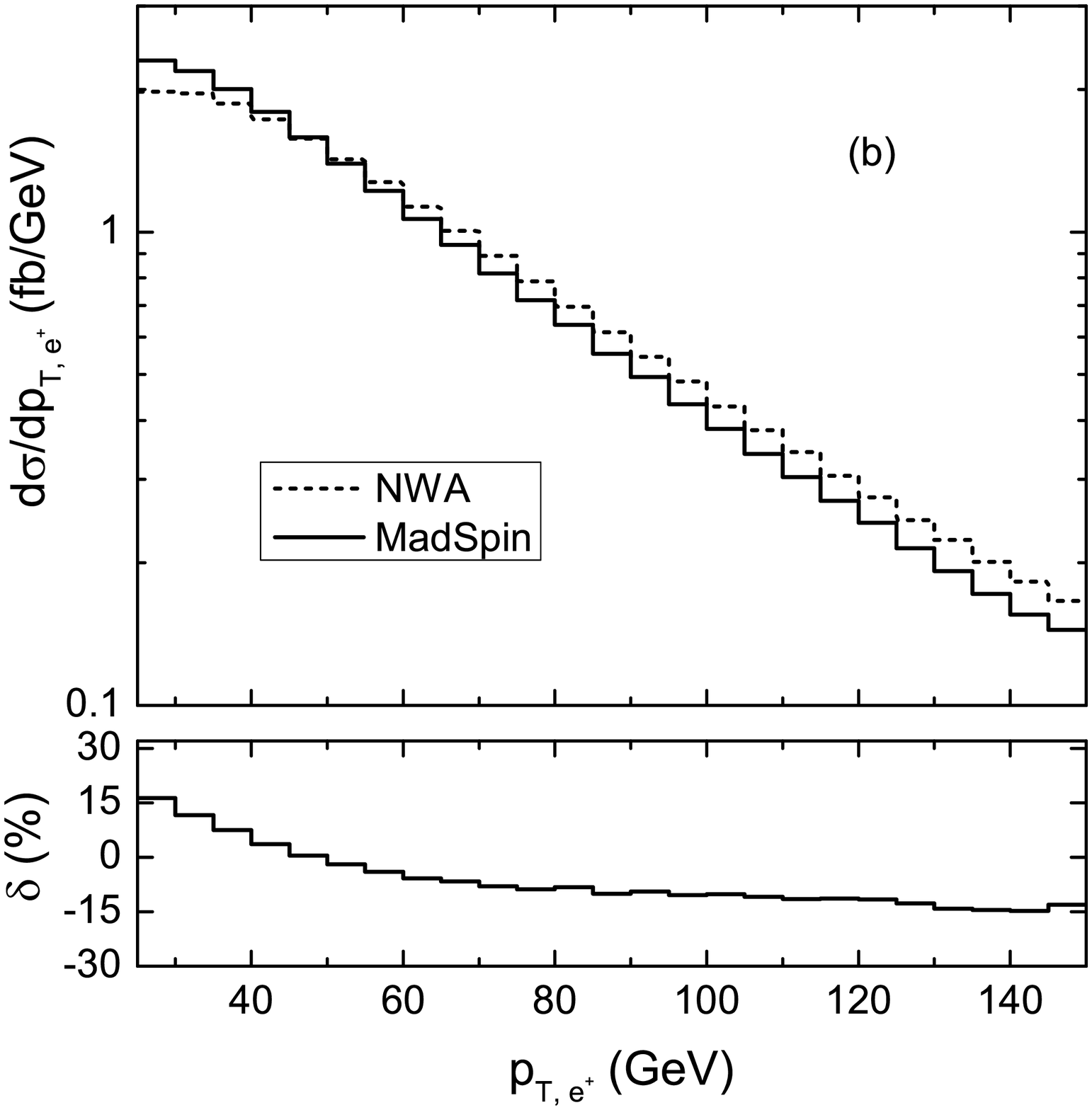}
\caption{\small The LO transverse momentum distributions of $e^+$ and the corresponding relative deviations by applying the naive NWA and {\sc MadSpin} methods with (a) $p_{T,j} > 50~{\rm GeV}$, and (b) $p_{T,j} >  50~{\rm GeV}$ and the additional constraints in (\ref{eqn:finalcut1}). }
\label{fig:loept}
\end{figure}

In Figs.\ref{fig:loept}(a,b) we depict the LO transverse momentum distributions of $e^+$ and the corresponding relative deviations by applying the naive NWA and {\sc MadSpin} methods. The results in Fig.\ref{fig:loept}(a) are obtained with only the constraint of $p_{T,j} > 50~{\rm GeV}$ on the leading jet. It shows that the spin correlation effect exists distinctly in the $p_{T,e^+}$ distribution, and the relative deviation varies from $-17.3\%$ to $24.9\%$ in the plotted $p_{T,e^+}$ region. In Fig.\ref{fig:loept}(b) both the cut of $p_{T,j} > 50~{\rm GeV}$ and the additional constraints in Eq.(\ref{eqn:finalcut1}) are taken in the event selection. We can see from the figure that the relative deviation between the {\sc MadSpin} and the naive NWA predictions changes from $-14.8\%$ to $16.3\%$ in the $p_{T,e^+}$ range of $[25,~150]~{\rm GeV}$, and reaches its maximum at the position of $p_{T,e^+} = 25~{\rm GeV}$.

\par
\begin{figure}[ht!]
\centering
\includegraphics[scale=0.35]{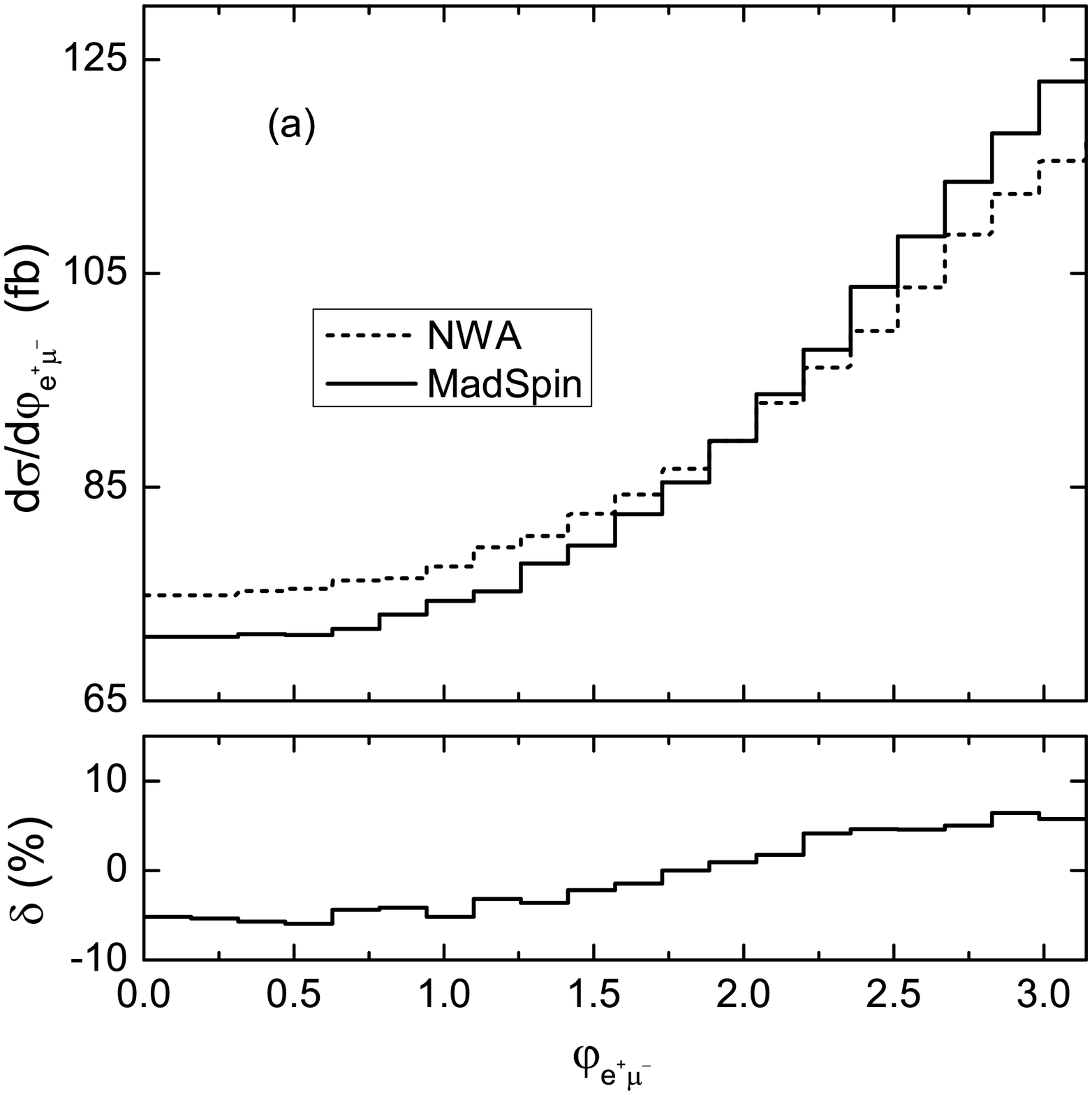}
\includegraphics[scale=0.35]{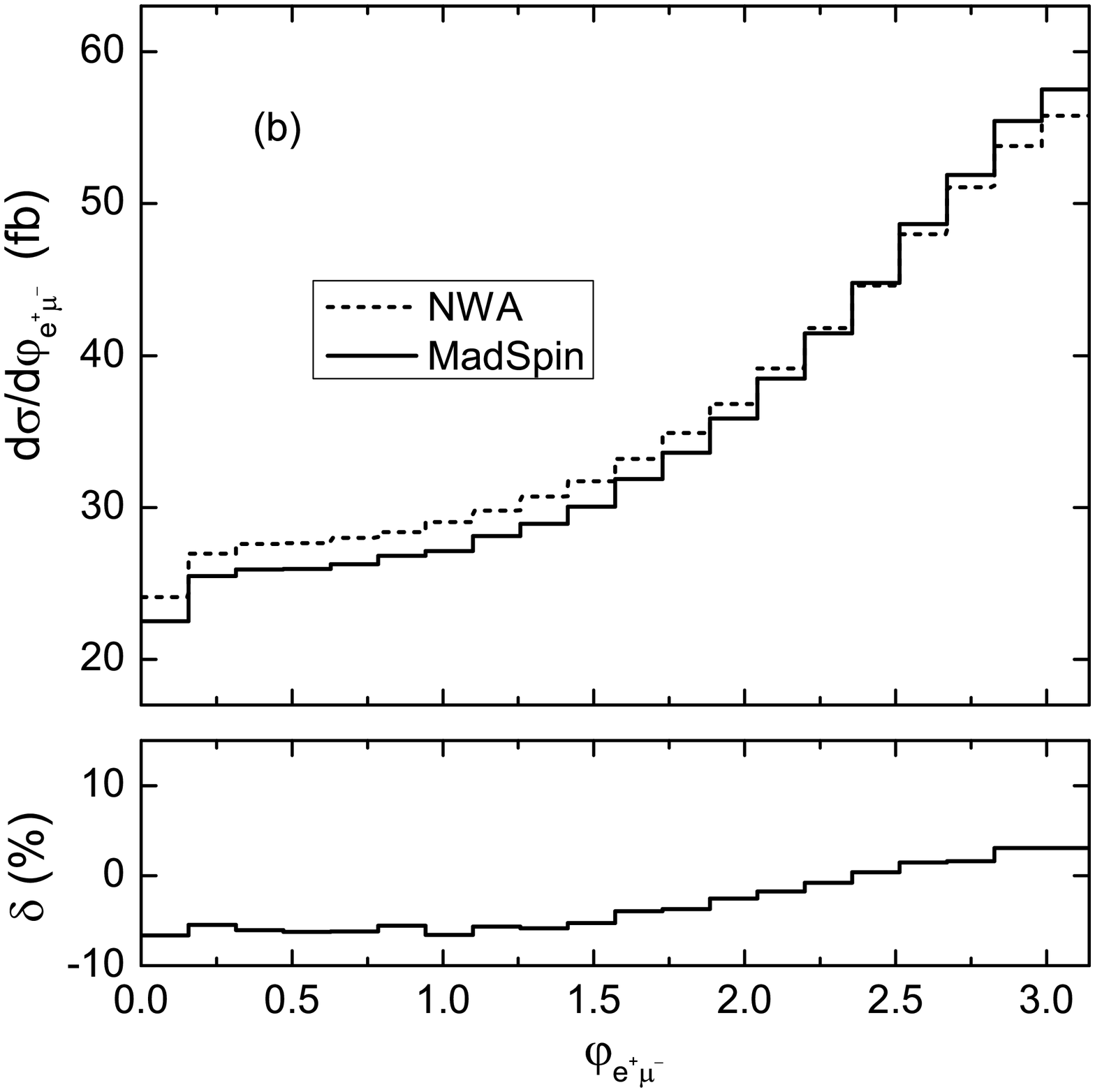}
\caption{\small The LO distributions of the azimuthal-angle difference $\phi_{e^+ \mu^-}$ between the final $e^+$ and $\mu^-$ and the corresponding relative deviations by applying the naive NWA and {\sc MadSpin} methods with (a) $p_{T,j} > 50~{\rm GeV}$, and (b) $p_{T,j} >  50~{\rm GeV}$ and the additional constraints in (\ref{eqn:finalcut1}).}
\label{fig:lophi}
\end{figure}

Figs.\ref{fig:lophi}(a,b) demonstrate the LO distributions of the azimuthal-angle difference $\phi_{e^+ \mu^-}$ between the final $e^+$ and $\mu^-$ on the transverse plane and the corresponding relative deviations by adopting the naive NWA and {\sc MadSpin} methods. Figs.\ref{fig:lophi}(a) and (b) are obtained by taking the same event selection criteria as in Figs.\ref{fig:loept}(a) and (b), respectively. We find that the $\phi_{e^+ \mu^-}$ distribution from the {\sc MadSpin} method has sizable deviation compared with that from the naive NWA method, and the relative deviation increases with the increment of $\phi_{e^+ \mu^-}$. In Fig.\ref{fig:lophi}(a) the relative deviation between the {\sc MadSpin} and the NWA predictions is in the range of $[-6.0\%,~6.4\%]$. Fig.\ref{fig:lophi}(b) demonstrates that $\delta(\phi_{e^+ \mu^-})$ varies from $-6.6\%$ to $3.1\%$ in the plotted range, and its absolute value can exceed $5\%$ in the region of $\phi_{e^+ \mu^-} < 1.5$.

\par
In Figs.\ref{fig:lomupt}(a,b), Figs.\ref{fig:loept}(a,b) and Figs.\ref{fig:lophi}(a,b) we see clearly that the spin correlation effect is demonstrated obviously in the LO $p_{T,\mu^-}$, $p_{T,e^+}$ and $\phi_{e^+ \mu^-}$ distributions, and the {\sc MadSpin} program is an efficient tool in handling the spin-entangled decays of resonant $W$-boson in an efficient and accurate way. In further investigations at the QCD NLO and EW NLO, we adopt only the {\sc MadSpin} method in event generation for the $pp \to W^+W^-+{\rm jet} \to e^+\mu^-\nu_e\bar{\nu}_{\mu}+{\rm jet}+X$ process.

\par
\subsubsection{NLO QCD and NLO EW corrected distributions}
\label{sec:nlodis}
\par
In this subsection, we discuss the NLO QCD and NLO EW corrected differential cross sections for the $pp \rightarrow W^+W^- + {\rm jet} + X$ process with the transverse momentum constraint of $p_{T,j} > 50~{\rm GeV}$ on the leading jet. For the distributions of the subsequent $W$-boson decay products, the additional constraints in Eq.(\ref{eqn:finalcut1}) are also applied in the event selection. The relative NLO QCD and NLO EW corrections to the differential cross section $d \sigma/dx$ are defined as
\begin{eqnarray}
  \delta^{{\rm QCD}}(x) = \biggl(\frac{d\sigma^{{\rm QCD}}_{{\rm NLO}}}{dx} - \frac{d\sigma_{{\rm LO}}}{dx}\biggr)\biggl/\frac{d\sigma_{{\rm LO}}}{dx} \, ,~~~~~~
  \delta^{{\rm EW}}(x) = \biggl(\frac{d\sigma^{{\rm EW}}_{{\rm NLO}}}{dx} - \frac{d\sigma_{{\rm LO}}}{dx}\biggr)\biggl/\frac{d\sigma_{{\rm LO}}}{dx}\, .
\end{eqnarray}

\par
\begin{figure}[ht!]
\centering
\includegraphics[scale=0.35]{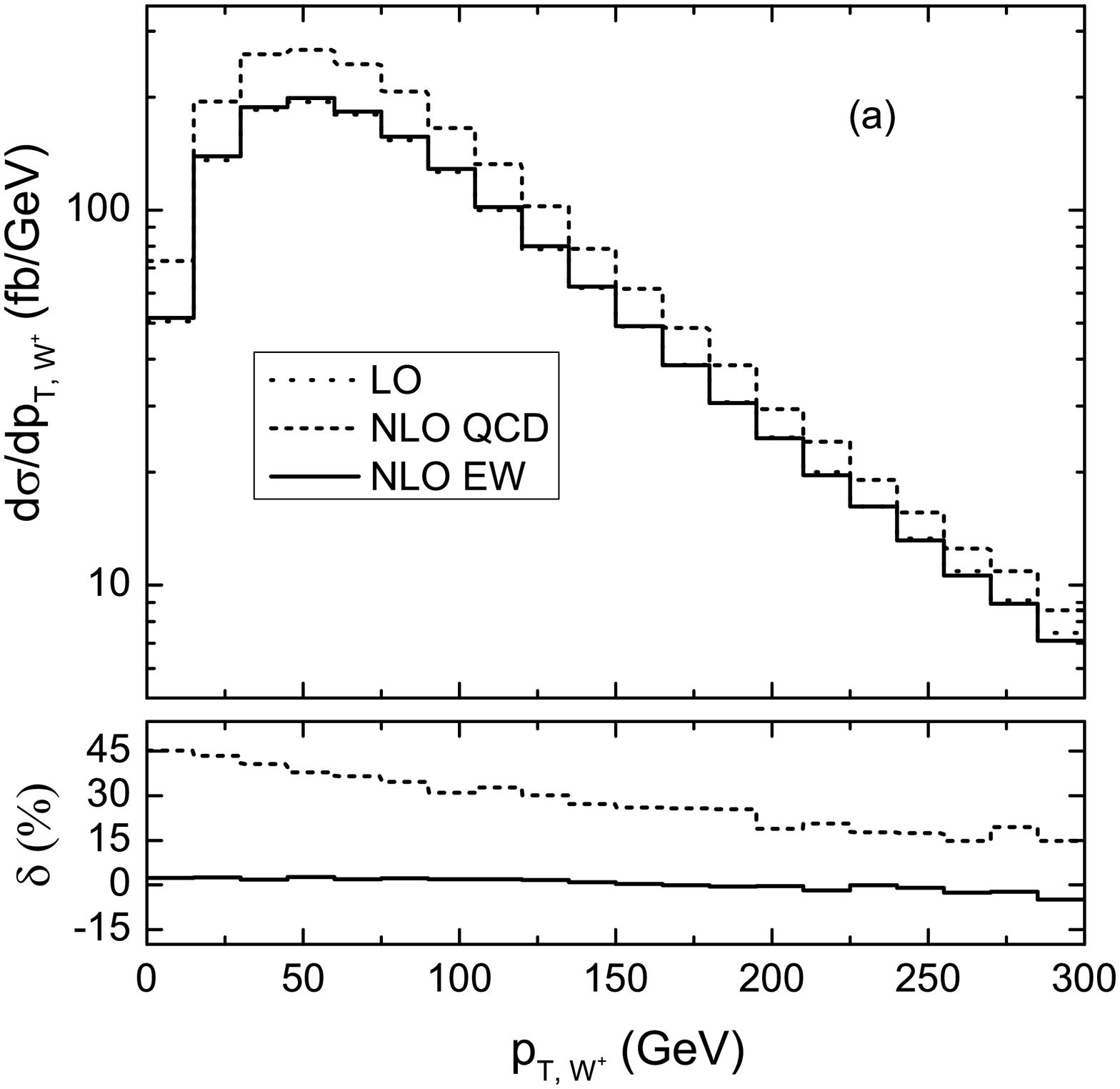}
\includegraphics[scale=0.35]{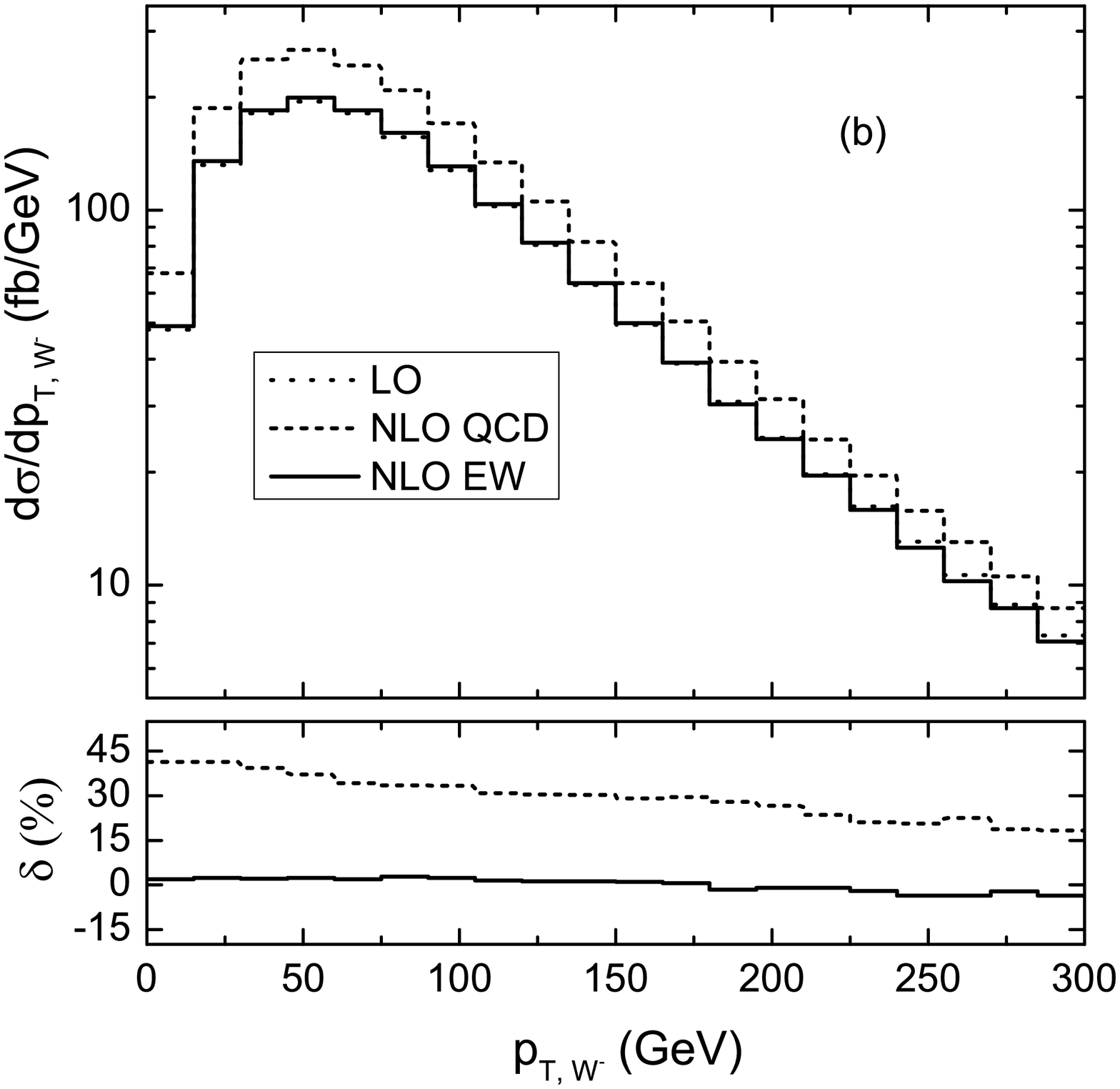}
\caption{\small The LO, NLO QCD and NLO EW corrected transverse momentum distributions of (a) $W^+$ and (b) $W^-$. }
\label{fig:nloWWpt}
\end{figure}

Figs.\ref{fig:nloWWpt}(a) and (b) describe the LO, NLO QCD and NLO EW corrected transverse momentum distributions of $W^+$ and $W^-$, respectively. The relative QCD corrections to the $p_{T,W^+}$ and $p_{T,W^-}$ distributions decrease from $45.2\%$ to $14.8\%$ and from $41.4\%$ to $18.3\%$ with the increment of $p_{T,W^{\pm}}$ from $0$ to $300~{\rm GeV}$, respectively. The relative EW corrections to the $p_{T,W^{\pm}}$ distributions decrease with the increment of $p_{T,W^{\pm}}$, and turn to be negative when $p_{T,W^{\pm}} > 180~{\rm GeV}$. Due to the large EW Sudakov logarithms, the EW correction becomes significant in high $p_{T,W^{\pm}}$ region. The relative EW corrections to the $p_{T,W^+}$ and $p_{T,W^-}$ distributions are about $-4.9\%$ and $-3.6\%$, respectively, at the position of $p_{T,W^\pm} = 300~{\rm GeV}$.

\par
\begin{figure}[ht!]
\centering
\includegraphics[scale=0.35]{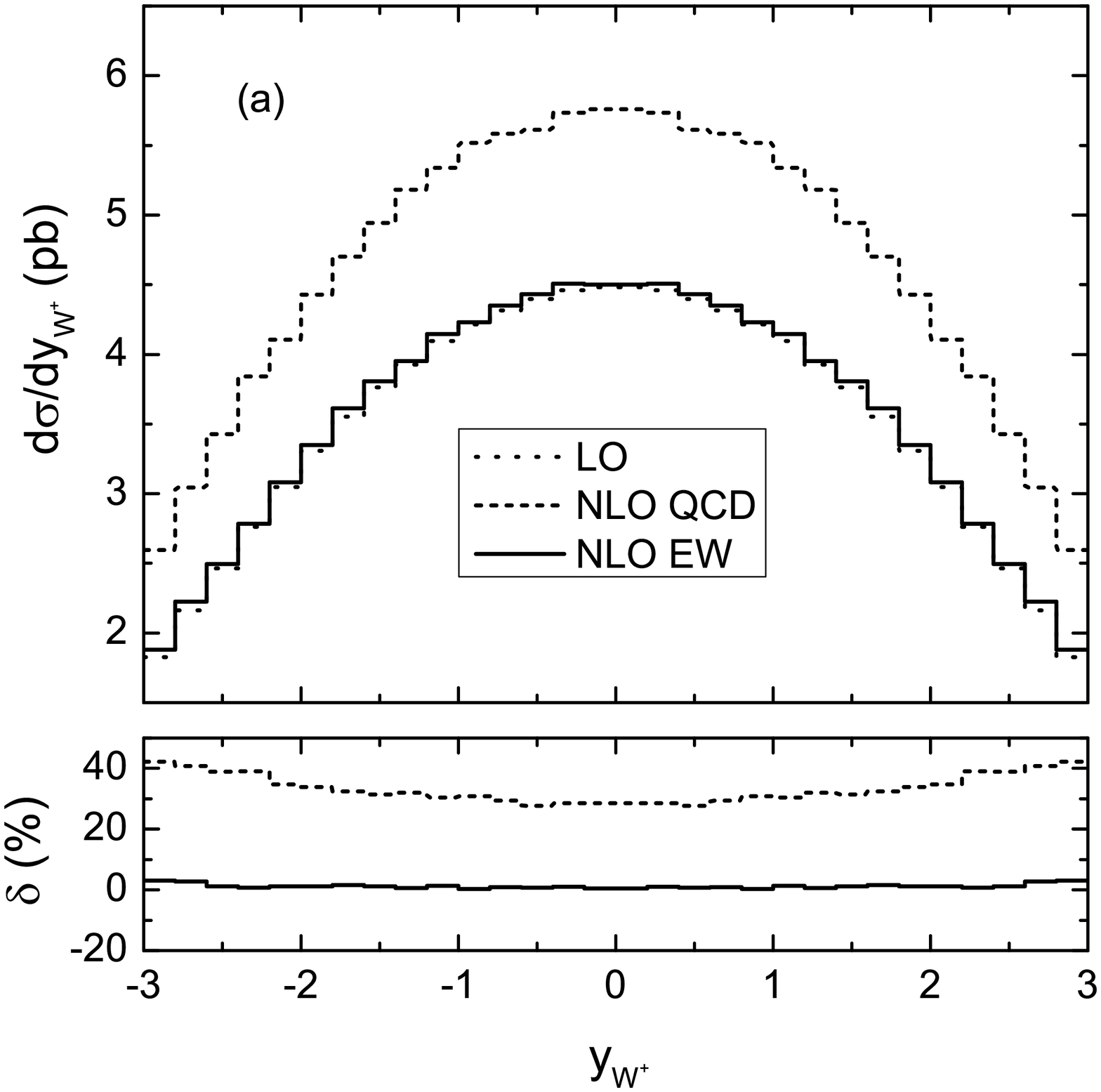}
\includegraphics[scale=0.35]{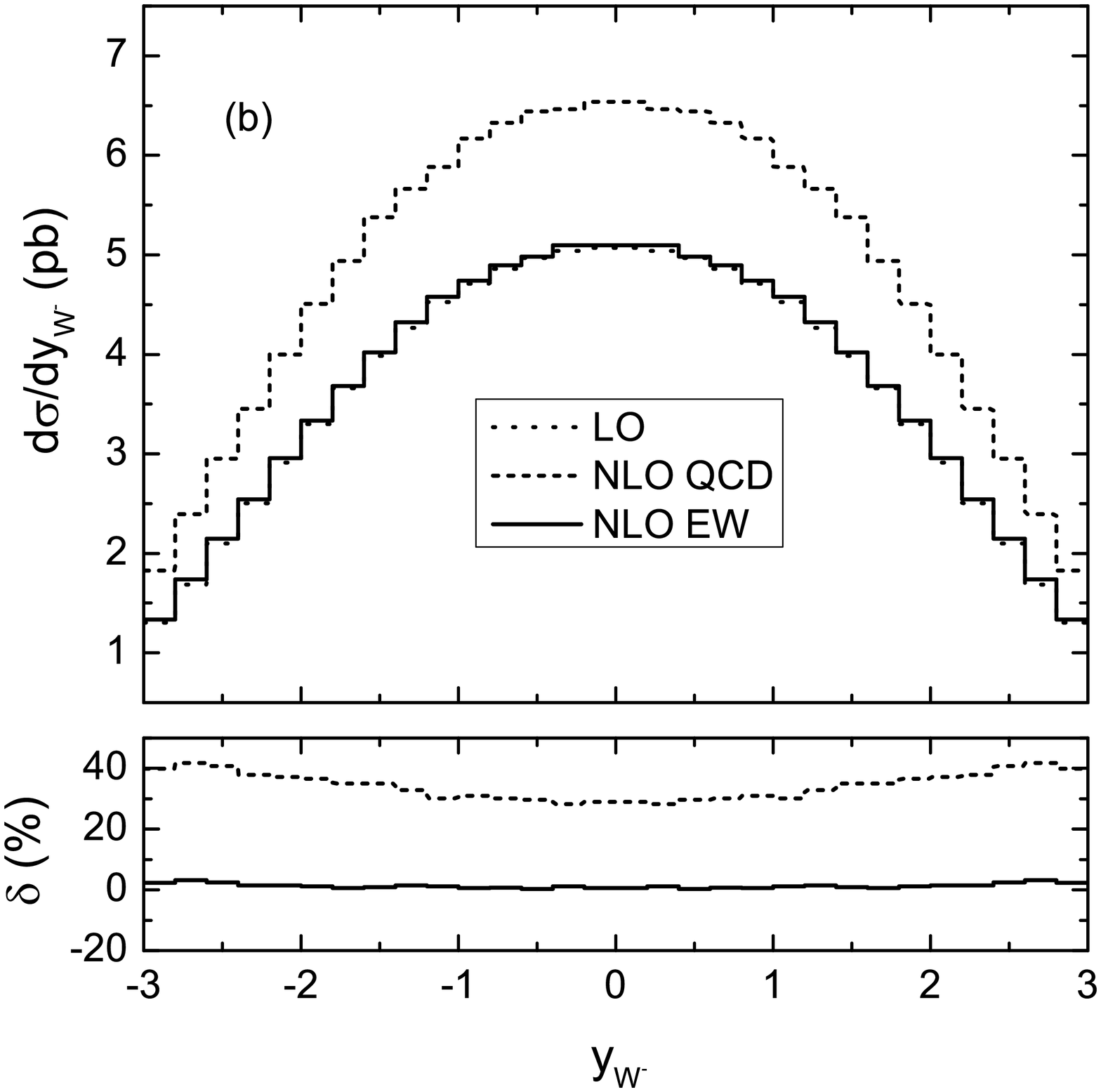}
\caption{\small The LO, NLO QCD and NLO EW corrected rapidity distributions of (a) $W^+$ and (b) $W^-$.}
\label{fig:nloWWrap}
\end{figure}

In Figs.\ref{fig:nloWWrap}(a) and (b) we present the LO, NLO QCD and NLO EW corrected rapidity distributions of $W^+$ and $W^-$ separately. We can see that the rapidity distributions of $W^-$-boson exhibit similar behavior with $W^+$. The relative QCD corrections to the $y_{W^+}$ and $y_{W^-}$ distributions increase from $27.6\%$ to $42.2\%$ and from $28.3\%$ to $41.9\%$, with the increment of $|y_{W^{\pm}}|$ from $0$ to $3$. For both $y_{W^+}$ and $y_{W^-}$ distributions, the relative EW corrections are less than $3\%$ in the plotted rapidity range.

\par
\begin{figure}[ht!]
\centering
\includegraphics[scale=0.35]{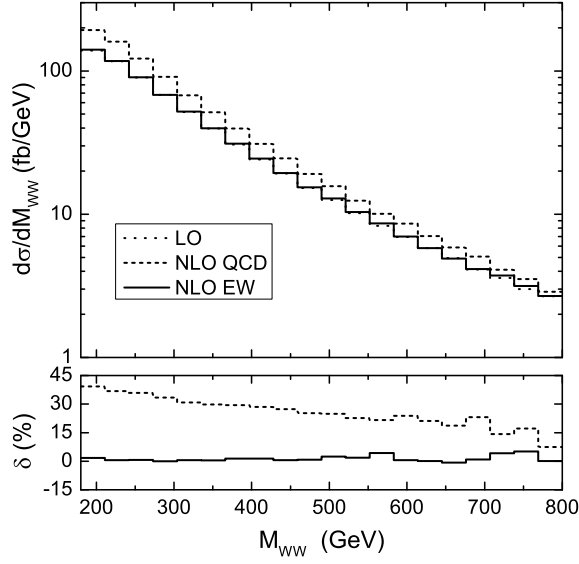}
\caption{\small The LO, NLO QCD and NLO EW corrected $W$-pair invariant mass distributions. }
\label{fig:nlomww}
\end{figure}

In Fig.\ref{fig:nlomww} we depict the LO, NLO QCD and NLO EW corrected $W$-pair invariant mass distributions and the corresponding relative corrections. With the increment of $M_{WW}$ from $180$ to $800~{\rm GeV}$, the relative QCD correction decreases from $39.2\%$ to $7.4\%$. The relative EW correction is always positive in the plotted range and can reach the maximal value of about $5\%$ due to the large positive contributions of the subprocesses with initial photon.

\par
\begin{figure}[ht!]
\centering
\includegraphics[scale=0.35]{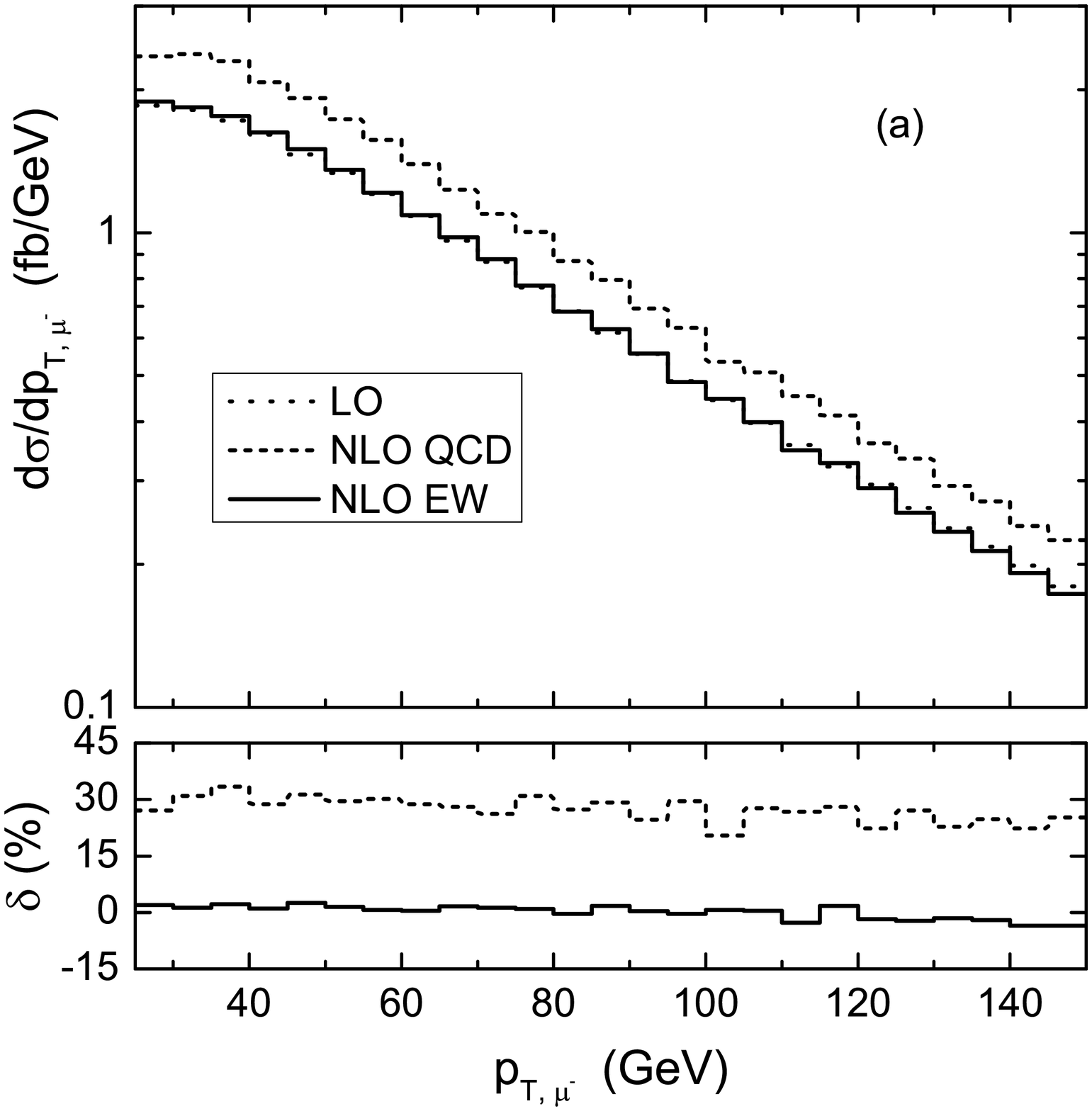}
\includegraphics[scale=0.35]{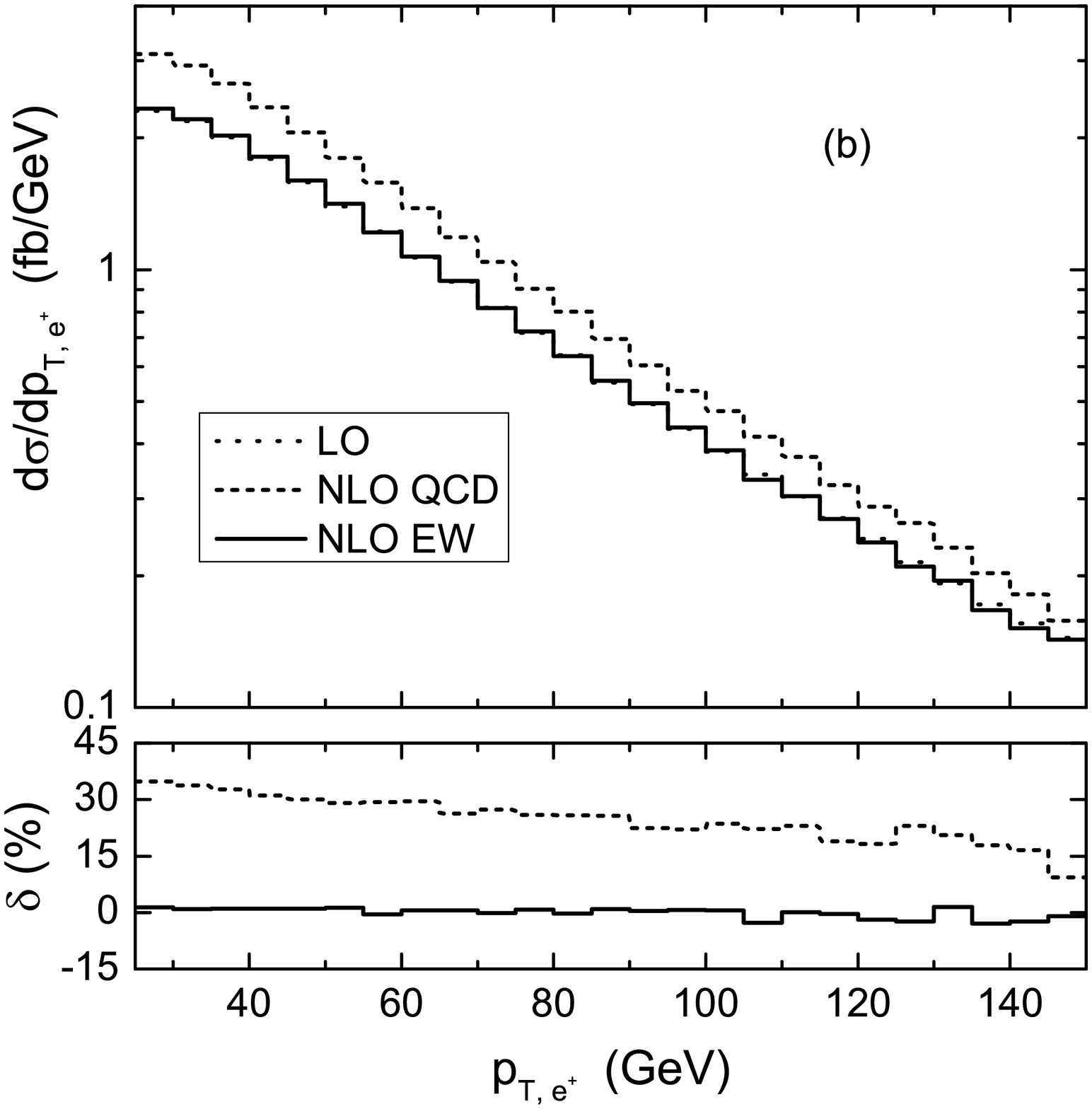}
\caption{\small The LO, NLO QCD and NLO EW corrected transverse momentum distributions of (a) $\mu^-$ and (b) $e^+$. }
\label{fig:nloleptonpt}
\end{figure}

The LO, NLO QCD and NLO EW corrected transverse momentum distributions of $\mu^-$ and $e^+$ are shown in Figs.\ref{fig:nloleptonpt}(a) and (b) separately. From the two figures we see that the relative QCD corrections to the $p_{T,\mu^-}$ and $p_{T,e^+}$ distributions vary in the ranges of $[22.3\%,~33.4\%]$ and $[9.3\%,~34.8\%]$, respectively, as $p_{T,\mu^-}$ and $p_{T,e^+}$ range from $25$ to $150~ {\rm GeV}$. At the position of $p_T = 150~{\rm GeV}$, the relative EW corrections to the $p_{T,\mu^-}$ and $p_{T,e^+}$ distributions are about $-3.5\%$ and $-1.0\%$, respectively.

\par
\begin{figure}[ht!]
\centering
\includegraphics[scale=0.4]{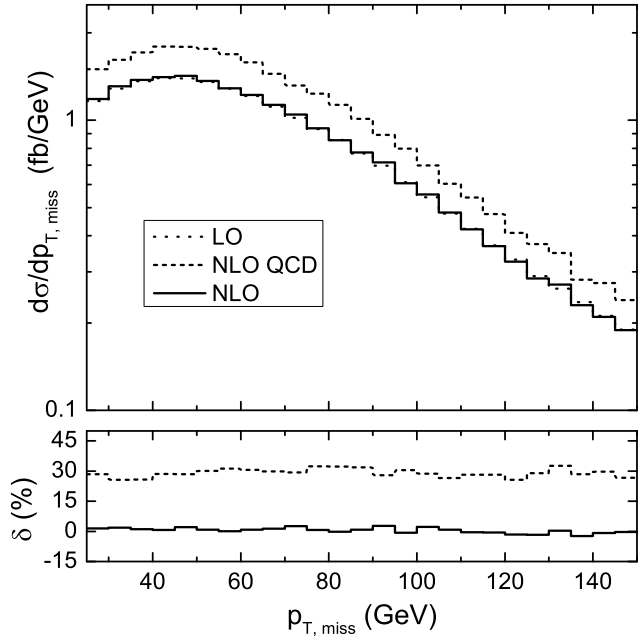}
\caption{\small The LO, NLO QCD and NLO EW corrected distributions of the missing transverse momentum.}
\label{fig:nloptmiss}
\end{figure}

In Fig.\ref{fig:nloptmiss}, we present the LO, NLO QCD and NLO EW corrected distributions of the missing transverse momentum. We find that the relative QCD correction to the $p_{T,{\rm miss}}$ distribution is stable and varies form $25.6\%$ to $32.7\%$ for $p_{T,{\rm miss}} \in [25,~150]~{\rm GeV}$. The relative EW correction to the $p_{T,{\rm miss}}$ distribution varies in the range of $[-2.3\%,~2.7\%]$, and reach its maximum at the position of $p_{T,{\rm miss}} \sim 90~{\rm GeV}$.

\vskip 5mm
\section{Summary}
\label{sec:conclusions}
\par
In this work we present the full NLO QCD and NLO EW corrections to the $W^+W^-+{\rm jet}$ production at the $14~ {\rm TeV}$ LHC followed by the $W$-boson leptonic decays. Our results show that the NLO EW correction is significant in high energy region due to the EW Sudakov effect which can be most probably detected in the LHC experiments. For example, with the jet transverse momentum constraint of $p_{T,j} > 1~{\rm TeV}$ on the leading jet, the NLO EW correction to the production rate for $pp \to W^+W^-+{\rm jet}+X$ can reach about $-22.2\%$ by taking $\mu=M_W$. In the calculations of the $pp \to W^+W^-+{\rm jet} \to e^+\mu^-\nu_e\bar{\nu}_{\mu}+{\rm jet}+X$ process, we adopt the {\sc MadSpin} method to include the spin correlation effect and find that the results by using the {\sc MadSpin} program are more accurate than using the naive NWA method. Therefore, we conclude that the spin correlation effect should be taken into account in precision calculations. We also present the LO, NLO QCD and NLO EW corrected distributions of final $W^\pm$-bosons and subsequent decay products, and find that the NLO EW correction is significant in high $W^\pm$-boson transverse momentum and high $W$-pair invariant mass regions.

\vskip 5mm
\section{Acknowledgments}
This work was supported in part by the National Natural Science Foundation of China (Grants No.11275190, No.11375008, No.11375171, and No.11405173), and the Fundamental Research Funds for the Central Universities (Grant No.WK2030040044).

\end{document}